\documentclass[prl,twocolumn,showpacs,preprintnumbers,amsmath,amssymb]{revtex4}

\usepackage{graphicx}
\usepackage{dcolumn}
\usepackage{bm}

\begin{document}

\preprint{APS/123-QED}

\title{Formation of bright matter-wave solitons during\\the collapse of Bose--Einstein condensates}

\author{Simon L. Cornish}
\email[Electronic address: ]{s.l.cornish@durham.ac.uk} \affiliation{ Department of Physics, Durham University, Rochester Building, Science
Laboratories, South Road, Durham DH1 3LE, United Kingdom.}
\author{Sarah T. Thompson}
\author{Carl E. Wieman}
\affiliation{JILA,
  National Institute of Standards and Technology and the University of
  Colorado, and the
  Department of Physics, University of Colorado, Boulder, Colorado
  80309-0440, USA.}

\date{\today}

\begin{abstract}
We observe bright matter-wave solitons form during the collapse of $^{85}\mbox{Rb}$ condensates in a three-dimensional magnetic
trap.  The collapse is induced by using a Feshbach resonance to suddenly switch the atomic interactions from repulsive to attractive. Remnant
condensates containing several times the critical number of atoms for the onset of instability are observed to survive the collapse. Under these
conditions a highly robust configuration of solitons forms such that each soliton satisfies the condition for stability and neighboring solitons
exhibit repulsive interactions.
\end{abstract}


\maketitle

Bose--Einstein condensates (BECs) with tunable interatomic interactions have been the subject of intense theoretical and experimental interest
in recent years.  Many aspects of these systems are now well understood. However, several questions remain pertaining to the response of the BEC
to a sudden change in the nature of the interactions from repulsive to attractive. Resolving such questions is critical if a complete
understanding of these basic atomic systems is to be developed and if related, more complex phenomena, such as the formation of molecular BECs and
superfluidity in degenerate Fermi systems, are to be thoroughly explained.

The macroscopic wave function of an atomic BEC obeys a nonlinear Schr\"{o}dinger equation, known as the Gross-Pitaevskii (GP) equation.  The
nonlinearity results from the interatomic interactions whose nature and magnitude are described by the s--wave scattering length ($a$).
Homogeneous condensates in free space are inherently unstable when the interactions are attractive ($a<0)$. However, in a harmonic trapping
potential, the zero--point kinetic energy can stabilize the condensate provided the number of atoms in the BEC $(N_0)$ is less than a critical
value $(N_{critical})$:
\begin{equation}\label{Ncritical}
    N_0<N_{critical}=k\frac{a_{ho}}{|a|}
\end{equation}
where $a_{ho}$ is the mean harmonic oscillator length which characterizes the kinetic energy in the trap, and $k$ is a dimensionless constant
known as the stability coefficient. The exact value of $k$ depends on the ratio of the trap frequencies \cite{Gammal2001a} and for our trap we
have measured it to be $k=0.46(6)$ \cite{Roberts2001a}.

In the vicinity of a Feshbach resonance, $a$ depends sensitively on the magnitude of an externally applied magnetic field
\cite{Stwalley1976,Tiesinga1992a,Tiesinga1993a}, allowing the magnitude and sign of the atom-atom interactions to be tuned to any value. We have
previously exploited such a resonance in $^{85}\mbox{Rb}$ to investigate the collapse of the condensate following a sudden change in the nature
of the interactions from repulsive to attractive \cite{Donley2001a}.  The resulting dynamics of the imploding ``Bosenova" were both fascinating
and unexpected. Most notable was our observation of an anisotropic burst of atoms that explodes from the condensate during the early stages of
collapse, leaving behind a highly excited remnant condensate that survives for many seconds. Curiously, we frequently observed that the number
of atoms in the remnant condensate, $N_{remnant}$, was significantly greater than $N_{critical}$.  It wasn't understood why this remnant BEC
didn't undergo further collapse until the number of atoms remaining was less than the value of $N_{critical}$ predicted by the GP equation. It
was particularly puzzling that the stability condition in Eq.\,\ref{Ncritical} accurately predicted the $onset$ of the collapse, but did not
seem consistent with number of atoms in the remnant.  In the work presented here, we explain this result by showing that the remnant is composed
of multiple solitons.

Solitons are localized waves that propagate over long distances without change in shape or attenuation. The existence of solitonic solutions is
a common feature of nonlinear wave equations. Solitons are therefore pervasive in nature, appearing in many diverse physical systems. BEC
systems described by the nonlinear GP equation can support both dark (local minima in the condensate wave function) and bright (local maxima)
solitons depending on whether the interactions are repulsive ($a>0$) or attractive ($a<0$), respectively. Dark solitons have been observed in
repulsive $^{87}\mbox{Rb}$ condensates \cite{Denschlag2000a,Burger1999a}, while bright solitons have been created using one-dimensional (1D)
$^7\mbox{Li}$ condensates with attractive interactions \cite{Strecker2002a,Khaykovich2002a}. Bright matter wave solitons form when the
attractive atomic interactions exactly balance the wave packet dispersion. Gap bright solitons have also been observed using $^{87}\mbox{Rb}$
condensates confined in an optical lattice \cite{Eiermann2004a}. In this letter we show that during the collapse of the condensate described in
ref.~\cite{Donley2001a}, a robust configuration of multiple solitons forms in our three-dimensional (3-D) trap.  Comparison with simulations
shows that, remarkably, the solitons are created with set relative phases such that they repel each other and oscillate in the trapping
potential for many seconds without degradation.  Since the number of atoms in each soliton never exceeds $N_{critical}$ and the repulsive
solitons never overlap, the condition for condensate stability is never violated.

The experimental apparatus and the magnetic field ramps for probing the condensate collapse have been described in detail elsewhere
\cite{Cornish2000a,Donley2001a}.  Stable $^{85}\mbox{Rb}$ condensates were produced within the positive scattering length region of a broad
Feshbach resonance by radio-frequency evaporation in a magnetic trap. The frequencies of the cylindrically symmetric cigar-shaped trap were $\nu
_{radial}=17.3\,\mbox{Hz}$ and $\nu _{axial}=6.8\,\mbox{Hz}$. Condensates of up to $15000$ atoms were formed with a shot-to-shot standard
deviation of less than $600$ atoms. The non-condensate thermal fraction was less than $500$ atoms. After the evaporation at
$162.2\,\mbox{Gauss}$, the magnetic field was adiabatically increased to $165.5\,\mbox{G}$ where $a=+9\,a_0$ ($a_0=0.0529\,\mbox{nm}$ is the
Bohr radius). The collapse of the condensate was then induced by quickly switching the scattering length through zero to a variable negative
value $a_{collapse}$ by increasing the magnetic field in $0.1\,\mbox{ms}$. After a variable hold time $t_{evolve}$ at the negative scattering
length, one of two imaging procedures was used to interrogate the remnant condensate. In the first, the scattering length was increased to a
large positive value in order to use the large repulsive atomic interaction to expand the condensate prior to the standard rapid trap turnoff
followed by absorption imaging sequence \cite{Donley2001a}. This procedure ensured that the size of the condensate was well above the resolution
limit of the imaging system ($7\,\mu\mbox{m}$ full-width at half-maximum (FWHM)), allowing the number of atoms to be measured reliably. In the
second, the magnetic trap was turned off immediately and the condensate imaged with minimum delay ($\sim 2\,\mbox{ms}$), effectively probing the
size and shape of the remnant condensate without any expansion.

\begin{figure}
\resizebox{7.5cm}{!}{\includegraphics[clip]{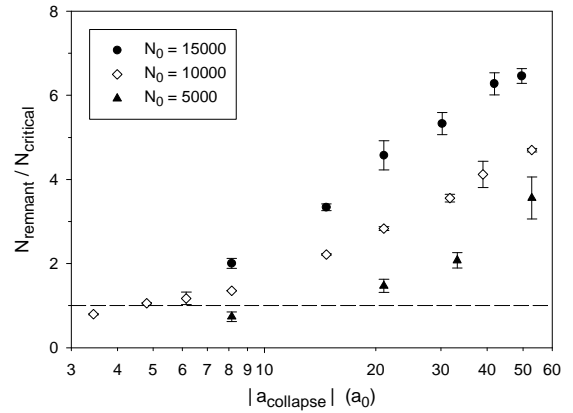}} \caption{\label{fig:Nremnant} Dependence of the number of condensate atoms surviving the
collapse on the magnitude of the negative scattering length at the collapse field ($a_{collapse}$ measured in units of the Bohr radius, $a_0 =
0.0529\,\mbox{nm}$). The data are expressed as the ratio of the number of remnant condensate atoms to the critical number calculated according
to Eq.\,\ref{Ncritical} for each scattering length. Results are shown for three different initial numbers of condensate atoms; $N_0=15000$
(\large$\bullet$\small), 10000 (\large$\diamond$\small) and 5000 (\small$\blacktriangle$\small). The dashed line shows the critical number. The
error bars represent the statistical spread in the data only.}
\end{figure}

The number of condensate atoms surviving the collapse was investigated as a function of both $a_{collapse}$ and the initial population of the
condensate. The hold time at $a_{collapse}$ was sufficiently long $(t_{evolve}=50\,\mbox{ms})$ to ensure that all collapse-related atom loss had
ceased \cite{Donley2001a}. The results are plotted in Fig.\,\ref{fig:Nremnant} as the ratio of the remnant number to the critical number
(calculated according to Eq.\,\ref{Ncritical} for each scattering length).  Depending upon the conditions of the collapse, $N_{remnant}$ varied
from below the critical number to much greater than the critical number (for $a_{collapse}=-100\,a_0$, the remnant condensate contained
approximately $10\times$ the critical number of atoms). Interestingly, the $fraction$ of atoms surviving the collapse depends only on the
scattering length and is independent of the initial number of condensate atoms (in the range we investigated). This fraction varies smoothly
from $\sim60\%$ for $a_{collapse}\simeq-5\,a_0$ to $\sim30\%$ for $a_{collapse}\simeq-50\,a_0$. The lifetime of the remnant condensate at
$a_{collapse}$ was several seconds and comparable to the lifetime of a condensate for small positive scattering lengths, even for
$N_{remnant}>N_{critical}$.

\begin{figure}
\resizebox{7.5cm}{!}{\includegraphics[clip]{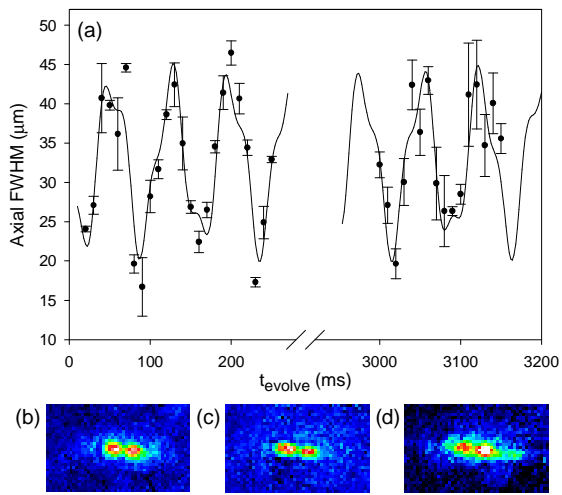}} \caption{\label{Fig:oscillation}(color online) Observation of solitons oscillating in the
magnetic trap following the collapse at a scattering length of $a_{collapse}=-11.4\,a_0$ of condensates initially containing approximately 8000
atoms. (a) The evolution of the axial (horizontal) FWHM of the remnant condensate obtained from a single Gaussian fit to the images. Above the
resolution limit of the imaging system, the remnant condensate is observed to separate into two solitons as shown in the images taken at (b)
$210\,\mbox{ms}$, (c) $1140\,\mbox{ms}$ and (d) $3110\,\mbox{ms}$. Each image is $77\times129\,\mu\mbox{m}$. The error bars represent the
statistical spread in the data only.}
\end{figure}

\begin{figure}[h]
\resizebox{7.5cm}{!}{\includegraphics[clip]{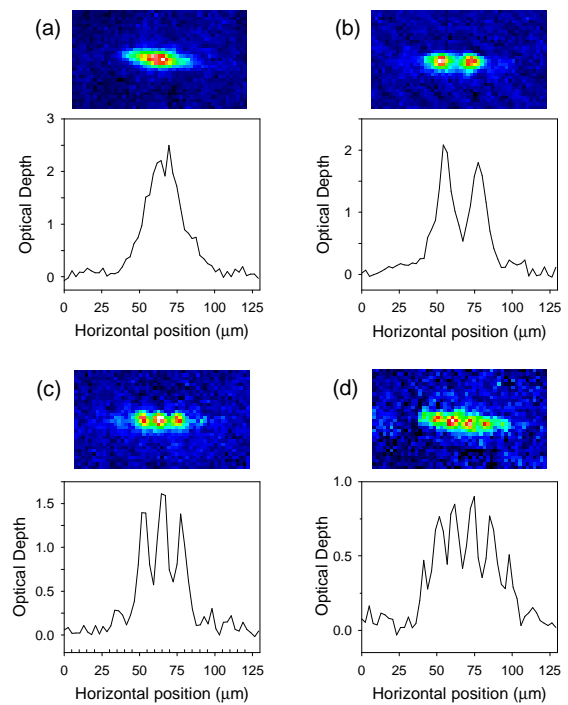}} \caption{\label{Fig:solitonimages} (color online) Images and cross-sections of remnant
condensates. (a) When the magnitude of $a_{collapse}$ is sufficiently small a single remnant condensate containing less than the critical number
is observed to survive the collapse. When the magnitude of $a_{collapse}$ is larger and/or larger initial condensates are used, the remnant
condensate is observed to split into a number of solitons determined by the conditions of the collapse (b-d). Each image is $77 \times
129\,\mu\mbox{m}$.}
\end{figure}

In a separate set of experiments, the evolution of the size and shape of the unexpanded remnant condensate was investigated. From the images,
the axial and radial column density distributions were calculated and fitted with Gaussian profiles to obtain both the axial and radial FWHM.
The result of this analysis is shown for the (weaker) axial direction in Fig.\,\ref{Fig:oscillation}\,(a). Any variation in the width of the
remnant condensate in the radial direction was below the resolution limit of our imaging system. The remnant condensate is highly excited, with
the variation in the axial width being well fitted by sinusoidal oscillations at the two lowest collective mode frequencies ($\nu_1=2\times
\nu_{axial}$ and $\nu_2=2\times \nu_{radial}$). The measured frequencies were $\nu_1=13.67(1)\,\mbox{Hz}$ and $\nu_2=33.44(3)\,\mbox{Hz}$. The
amplitude of the $\nu_2$ oscillation was $29(7)\%$ as large as that of the $\nu_1$ oscillation and contained a significant contribution from the
radial focus of the atom burst associated with the collapse \cite{Donley2001a}. There was no discernible damping during our 3~second observation
time.

Close inspection of the images, however, revealed that the observed variation in the width is not due to the excitation of collective modes. The
remnant condensate is observed to separate into two (or more) distinct clouds which we associate with solitons
(Fig.\,\ref{Fig:oscillation}\,(b-d)). The observed oscillation in the axial width of the remnant then results from the oscillation of two
solitons along the axial direction of the trap. As was observed for the soliton trains created in the $^7\mbox{Li}$ experiments
\cite{Strecker2002a}, neighboring solitons form with a relative phase, $\phi$, that ensures that they interact repulsively ($\pi/2<\phi<3\pi/2$)
even though the atomic interactions are attractive \cite{Al Khawaja2002a,Salasnich2002a}. In this case, the solitons do not pass through each
other at the center of the trap, but rather rebound off one another. Consequently, as the solitons never fully overlap, the critical density for
collapse is never reached and the solitons remain stable. This conclusion is supported by the observation that the soliton structures persist
for over three seconds (Fig.\,\ref{Fig:oscillation}\,(c)) with only a small degradation in the quality of the images due to the presence of more
thermal atoms as the condensate melts. The solitons collide over 40 times during this period. Several authors propose that bright matter wave
solitons form due to a modulation instability of the condensate wave function \cite{Carr2000a,Salasnich2003a,Carr2004a}. Our previous
observations of local `spikes' in the condensate density \cite{Donley2001a} are consistent with this mechanism.

\begin{figure}[t]
\resizebox{7.5cm}{!}{\includegraphics[clip]{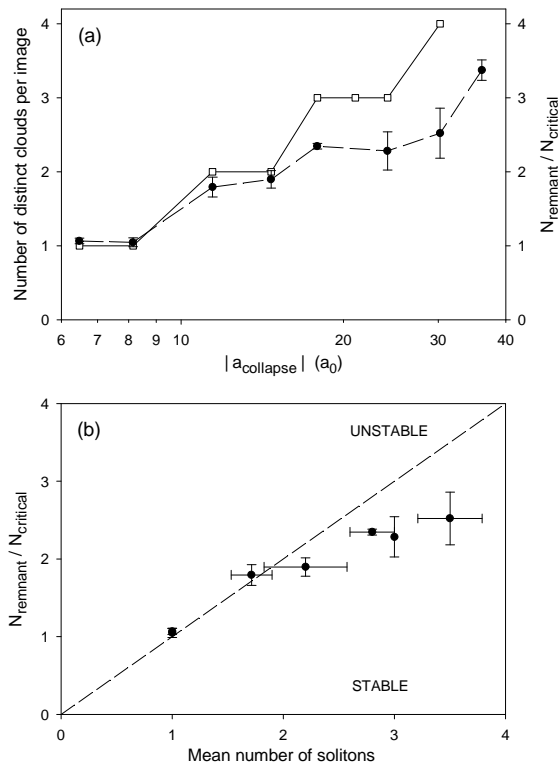}} \caption{\label{Fig:solitonnumber} (a) The modal number of solitons
(\scriptsize$\square$\small) observed after $200\,\mbox{ms}$ of evolution at the collapse field as a function of the magnitude of the negative
scattering ($a_{collapse}$). The ratio of the number of remnant condensate atoms to the critical number (\large$\bullet$\small) is also plotted
on the right axis (note the two y-axes have the same scale). The initial condensate contained approximately 8000 atoms. (b) The ratio of the
number of remnant condensate atoms to the critical number is plotted against the mean number of observed solitons to show that each soliton
contains less than the critical number of atoms defined by Eq.\,\ref{Ncritical} and is therefore expected to be stable. The error bars represent
the statistical spread in the data only.}
\end{figure}

The number of solitons that form depends both on the magnitude of $a_{collapse}$ and the initial number of condensate atoms. In
Fig.\,\ref{Fig:solitonimages} we present images and cross sections showing two (Fig.\,\ref{Fig:solitonimages}\,(b)) and three
(Fig.\,\ref{Fig:solitonimages}\,(c)) solitons. These images were acquired with $t_{evolve}=200\,\mbox{ms}$, when the axial width was a maximum
and the solitons were maximally separated. For comparison, Fig.\,\ref{Fig:solitonimages}\,(a) shows a remnant condensate for which the number of
remnant atoms was below the critical number. Under these conditions the remnant was never observed to separate into two or more solitons.  We
have observed up to six solitons present (Fig.\,\ref{Fig:solitonimages}\,(d)). However, the quality of the images was significantly reduced when
more than three solitons were present because the spacing between solitons became comparable to the imaging resolution limit. In
Fig.\,\ref{Fig:solitonimages}\,(c) the three solitons contain approximately the same number of atoms. This was not always the case, however, and
we often observed an unequal and (usually) symmetric population distribution across the solitons.

In Fig.\,\ref{Fig:solitonnumber} we examine in more detail the conditions under which multiple solitons form. For a fixed initial number of
atoms in the condensate ($N_0\approx 8000$), the number of solitons created was investigated as a function of $a_{collapse}$. The images were
again acquired with $t_{evolve}=200\,\mbox{ms}$. In Fig.\,\ref{Fig:solitonnumber}\,(a) the modal number of solitons is plotted. The data show a
gradual increase in the number of solitons as the magnitude of $a_{collapse}$ increases. The ratio of the number of remnant condensate atoms to
the critical number for the same $N_0$ is also plotted in Fig.\,\ref{Fig:solitonnumber}\,(a). This ratio exhibits the same general trend as a
function of $a_{collapse}$ as the observed number of solitons. Most importantly, it never exceeds the number of solitons. From this observation we
deduce that the number of atoms in each soliton is always less than or equal to the critical number. Each soliton is therefore stable against collapse,
explaining the unexpected stability of remnant condensates for which the number of atoms is greater than the critical number. This
point is further illustrated in Fig.\,\ref{Fig:solitonnumber}\,(b) where the ratio of the number of remnant condensate atoms to the critical
number is plotted against the observed mean number of solitons. All the points lie on or below the critical condition defined by
Eq.\,\ref{Ncritical} (dashed line).

In contrast to the previous observations of bright solitons in one-dimensional confining potentials \cite{Strecker2002a,Khaykovich2002a}, these
experiments were performed in a 3D trap with an aspect ratio of $\nu_{radial}/\nu_{axial}\approx2.5$. Under such conditions, 3D bright solitons
are predicted to be stable only if the attractive atomic interaction is smaller than a critical value \cite{Salasnich2002a}, consistent with our
observations. Simulations of our experiment based on a numerical integration of the 3D GP equation show that multiple solitons are only stable
provided they interact repulsively \cite{Parker2005a}. Attractive interactions between the solitons would lead to energetic excitations of the
soliton wave function during each `collision' (in this case the solitons pass through one another), and ultimately a loss of particles from the
BEC. Moreover, when the relative phase between solitons is fixed at $\pi$ in the simulation, the maximum number of solitons that is predicted to
be stable at a particular scattering length is in good agreement with our observations shown in Fig.\,\ref{Fig:solitonnumber}.

The sudden change in the nature of the atomic interactions in a Bose--Einstein condensate from repulsive to attractive leads to a violent
collapse process, in which a significant proportion of the initial condensate is ejected in a highly energetic burst. Yet, surprisingly, out of
this violent explosion emerges a robust configuration of solitons that preserves a large
fraction of the initial condensate population. It is particularly striking that this behavior holds over a large range of attractive interaction
strengths.

\begin{acknowledgments}
We thank N.\,Parker and C.\,S.\,Adams for helpful discussions. SLC acknowledges the support of the Royal Society. This work was supported by ONR
and NSF.
\end{acknowledgments}


\end{document}